\documentclass[floatfix,amssymb,prl,twocolumn,superscriptaddress,preprintnumbers,nofootinbib, aps]{revtex4-1}
		
\usepackage{amssymb,amsmath,verbatim,mathtools,needspace,enumitem,etoolbox,graphicx,physics,microtype,afterpage,bm,soul}
\usepackage[dvipsnames]{xcolor}
\definecolor{linkcolor}{rgb}{0.0,0.3,0.5}
\usepackage[unicode, colorlinks=true, linkcolor=linkcolor, citecolor=linkcolor, filecolor=linkcolor,urlcolor=linkcolor, pdfusetitle]{hyperref}
\usepackage{orcidlink}
\usepackage[all]{hypcap}
\usepackage[T1]{fontenc}
\usepackage[utf8]{inputenc}
\usepackage{tabularx}
\usepackage{float}
\usepackage{placeins}
\renewcommand{\arraystretch}{1.4}

\usepackage{lmodern}
\usepackage{ragged2e}
\allowdisplaybreaks
\usepackage{tikz}
\usepackage{color}
\usepackage{framed}
\usepackage{hyperref}
\hypersetup{colorlinks, citecolor=rossos, linkcolor=black, urlcolor=rossos}
\definecolor{rossos}{cmyk}{0,1,1,0.55}
\definecolor{bluscuro}{rgb}{0.15, 0.2, .85}
\definecolor{bluchiaro}{cmyk}{1,.3,0.,0.1}
\definecolor{ForestGreen}{rgb}{0.13, 0.55, 0.13}
\definecolor{darkblue}{rgb}{0,0, 1.39}

\newcommand{\be}{\begin{equation}}
\newcommand{\ee}{\end{equation}}

\def\BH{\text{\tiny BH}}

\newcommand{\PBH}{\text{\tiny PBH}}

\usepackage{xcolor}
\definecolor{darkgreen}{RGB}{0,100,0}

\usepackage{pifont}
\newcommand{\cmark}{\ding{51}}%
\newcommand{\xmark}{\ding{55}}%

\def\lsim{\mathrel{\rlap{\lower4pt\hbox{\hskip0.5pt$\sim$}}
    \raise1pt\hbox{$<$}}}         
\def\gsim{\mathrel{\rlap{\lower4pt\hbox{\hskip0.5pt$\sim$}}
    \raise1pt\hbox{$>$}}}         

\newcommand{\jhu}{William H.\ Miller III Department of Physics and Astronomy, Johns Hopkins University, \\ 3400 North Charles Street, Baltimore, Maryland, 21218, USA}
\newcommand{\cern}{Department of Theoretical Physics, CERN, Esplanade des Particules 1, P.O. Box 1211, Geneva 23, Switzerland}
\newcommand{\unipd}{Dipartimento di Fisica e Astronomia ``G. Galilei'', Università degli Studi di Padova, via Marzolo 8, I-35131 Padova, Italy}
\newcommand{\infnpd}{INFN, Sezione di Padova, via Marzolo 8, I-35131 Padova, Italy}

\begin{document}

\preprint{CERN-TH-2025-261}

\title{Primordial Black-Hole-Based Pathways to Little Red Dots}

\author{Valerio De Luca\orcidlink{0000-0002-1444-5372}}
\email{vdeluca2@jh.edu}
\affiliation{\jhu}

\author{Loris Del Grosso\orcidlink{0000-0002-6722-4629}}
\email{ldelgro1@jh.edu}
\affiliation{\jhu}

\author{Gabriele Franciolini\orcidlink{0000-0002-6892-9145}}
\email{gabriele.franciolini@unipd.it}
\affiliation{\unipd}
\affiliation{\infnpd}
\affiliation{\cern}

\author{Konstantinos Kritos\orcidlink{0000-0002-0212-3472}}
\email{kkritos1@jhu.edu}
\affiliation{\jhu}

\author{Emanuele Berti\orcidlink{0000-0003-0751-5130}}
\email{berti@jhu.edu}
\affiliation{\jhu}

\author{Daniel J. D'Orazio\orcidlink{0000-0002-1271-6247}}
\email{dorazio@stsci.edu}
\affiliation{Space Telescope Science Institute, 3700 San Martin Drive, Baltimore, Maryland 21218, USA}
\affiliation{\jhu}
\affiliation{Niels Bohr International Academy, Niels Bohr Institute, Blegdamsvej 17, DK-2100 Copenhagen Ø, Denmark}

\author{Joseph Silk\orcidlink{0000-0002-1566-8148}}
\email{silk@iap.fr}
\affiliation{\jhu}
\affiliation{Institut d'Astrophysique de Paris, UMR 7095 CNRS and UPMC, Sorbonne Université, F-75014 Paris, France}
\affiliation{Department of Physics, Beecroft Institute for Particle Astrophysics and Cosmology, University of Oxford, Oxford OX1 3RH, United
Kingdom}


\begin{abstract}
\medskip
\noindent
The James Webb Space Telescope has uncovered a population of compact, high-redshift sources, the Little Red Dots (LRDs), which may host supermassive black holes (BHs) significantly heavier than their stellar content compared with local scaling relations. These objects challenge standard models of early galaxy formation and may represent an extreme class of early BH hosts. In this Letter, we investigate whether these BHs could have a primordial origin. We first show that the direct formation of these BH masses in the early Universe is excluded by stringent cosmic microwave background $\mu$-distortion limits. We then investigate the assembly of massive BHs from lighter, observationally allowed primordial black holes (PBHs) via hierarchical mergers, finding that, although this channel can operate depending on the merger history, it faces challenges in explaining the observations due to the rarity of the required high-redshift dark matter halos. Finally, we estimate gas accretion onto intermediate-mass PBHs, while jointly tracking metallicity evolution, and identify regions of parameter space in which such growth could reproduce the observed properties of LRDs. As a special case, we focus on the strongly lensed source QSO1, whose extremely low metallicity and large mass provide a stringent test of these formation channels. 
\end{abstract}
\maketitle

\vspace{0.1cm}
\noindent{{\bf{\em Introduction.}}} 
Recent observations with the James Webb Space Telescope have revealed a population of compact, optically red sources at high redshifts, commonly referred to as LRDs~\cite{Pacucci:2023oci, Matthee:2023utn,2023ApJ...959...39H,Greene:2024phl, 2023ApJ...954L...4K, Akins:2025iqo,2025ApJ...986..165T,2025arXiv251203130I}. These objects are characterized by V-shaped rest-frame ultraviolet-optical continua, compact morphologies ($\lesssim100\,\rm pc$), and broad emission features, distinguishing them from typical star-forming galaxies or other previously discovered quasars in the early Universe~\cite{Greene:2024phl, Kocevski:2025tft, 2024arXiv241103424S,2025ApJ...978...92L, Baggen:2024vef, Baggen2025}. These characteristics strongly suggest that LRDs are associated with accreting BHs at early epochs. Yet, several of their properties remain difficult to reconcile with standard models of early active galactic nuclei (AGN)~\cite{Matthee:2023utn,2025A&A...702A..57H,2025ApJ...988L..22I}. The inferred masses of their putative BHs are on the order of $(10^{6} - 10^{8})M_\odot$, with unusually high BH-to-stellar mass ratios relative to the local $M_\BH - M_\star$ relation~\cite{Pacucci:2023oci, Matthee:2023utn,Greene:2024phl,Maiolino:2023bpi}. LRDs also appear to be abundant, with reported number densities approaching 10\% -- 20\% of the galaxy population for an absolute UV magnitude between ${\rm M}_\text{\tiny UV} \approx -19$ and  ${\rm M}_\text{\tiny UV} \approx -22$~\cite{Kocevski:2025tft, Matthee:2023utn, Maiolino:2023bpi}. Moreover, LRDs exhibit weak hot-dust emission and a conspicuous lack of detectable x-ray signatures, which together pose significant challenges to interpretations based solely on AGN templates~\cite{2024ApJ...974L..26Y,2025ApJ...991...37A,Ananna:2024jug,Maiolino:2024uon, 2025ApJ...994..113L, 2025A&A...693L...2P, Mazzolari:2024htb, 2024ApJ...968....4P, 2024ApJ...975L...4C,Kokubo:2024ukw}. These
non-detections of typical AGN-like features may suggest alternative explanations based on massive and compact
galaxies~\cite{2025ApJ...978...92L, 2023ApJ...956...61A,2025ApJ...991...37A}, hierarchical growth within dense environments such as nuclear star clusters~\cite{Kritos:2024upo,Kritos:2024sgd,Kritos:2025aqo}, supermassive stars~\cite{Begelman:2025upi,Zwick:2025eik, NandalLoeb:2025}, BH stars~\cite{Naidu:2025rpo} and  direct collapse BH models~\cite{Pacucci:2026ovn,Qin:2025ymc}. 
Notice that a modest fraction of the LRDs observed masses may also be attributed to tidal disruption events~\cite{Kritos:2025aqo}. It has also been suggested that super-Eddington accretion phases might help explain the lack of x-ray and variability signatures in cases where LRDs are actually AGNs~\cite{Pacucci:2024tws, Inayoshi:2024rav,Lambrides:2024ugh, Madau:2024fdv, 2025ApJ...994..113L}.

An illustrative example is the strongly lensed LRD source QSO1 at redshift $z=7.04$. Previous studies~\cite{Maiolino:2025tih} have reported that QSO1 shows exceptionally low metallicity ($Z \lesssim 0.01 \,Z_\odot$) and hosts an overmassive central BH, with mass~$\sim 5 \times 10^{7} M_\odot$ and an inferred ratio $M_\BH/M_\star > 2$, placing it well above the local $M_\BH - M_\star$ relation. More recent analysis~\cite{2025arXiv250821748J} has confirmed the BH mass through narrow-line resolved kinematics, strengthening the interpretation that QSO1 contains a rapidly growing BH seed within a compact and metal-poor environment. These properties make QSO1 a benchmark for understanding the diversity of LRDs and the mechanisms that drive extreme BH growth in the early Universe.

A possible interpretation for QSO1, and potentially for the broader population of LRDs, invokes PBHs as the initial seeds of their central massive objects~\cite{Zeldovich:1967lct,Hawking:1971ei,Carr:1974nx,Carr:1975qj} (see e.g.~\cite{Sasaki:2018dmp, Carr:2020gox, Green:2020jor,Byrnes:2025tji,LISACosmologyWorkingGroup:2023njw} for recent reviews).  These objects can form from the gravitational collapse of high-redshift large-amplitude perturbations, leading to a possibly wide range of PBH masses (see also~\cite{LISACosmologyWorkingGroup:2023njw,Byrnes:2025tji} for alternative formation scenarios). PBHs are of particular interest because they could also make a significant contribution to dark matter (DM), and potentially account for some of the binary BH mergers observed so far~\cite{Bird:2016dcv, Sasaki:2016jop, Clesse:2016vqa, Mukherjee:2021ags, DeLuca:2020sae, DeLuca:2020qqa, DeLuca:2020agl, Franciolini:2021tla, Afroz:2024fzp,Berti:2025usa} by the LIGO-Virgo-KAGRA Collaboration~\cite{LIGOScientific:2016aoc, LIGOScientific:2018mvr, LIGOScientific:2020ibl, KAGRA:2021vkt, LIGOScientific:2020iuh, LIGOScientific:2025rsn}.  

In this Letter, we examine whether QSO1 is consistent with a PBH origin and assess the implications for the broader population of LRDs. Previous studies~\cite{Zhang:2025asq,Zhang:2025grn, Maiolino:2025tih} suggested that a PBH scenario could naturally account for the low metallicity observed in high-redshift LRDs. Furthermore, PBH seeds provide a pathway that bypasses the formation of stellar-mass BH remnants from the first stars, enabling rapid growth into the observed LRD population through efficient gas accretion~\cite{Dayal:2024zwq,Dayal:2025aiv,Ziparo:2024nwh,Matteri:2025vnv}. Here, we explore three primordial-origin scenarios for the BHs powering LRDs: 
{\it (i)} direct PBH formation from high-redshift overdensities;
{\it (ii)} hierarchical assembly through PBH mergers;
{\it (iii)} growth of PBH seeds via gas accretion. 
We find that direct formation is effectively ruled out by existing constraints; hierarchical mergers face challenges in explaining the inferred number densities of LRDs, whereas accretion onto PBH seeds provides the most viable pathway.

Throughout we use natural units ($\hbar = c = 1$) and assume the $\Lambda$CDM cosmology~\cite{Planck:2018vyg}.

\begin{figure}
    \centering
    \includegraphics[width=0.99
    \linewidth]{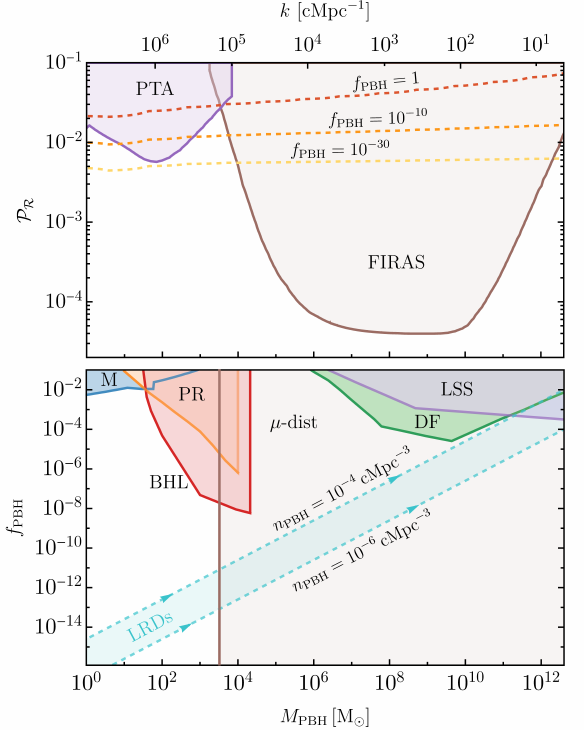}
    \caption{Constraints on the parameter space of PBH formation. Upper: limits on the amplitude of the primordial power spectrum, in terms of the high-redshift PBH mass $M_\PBH$. Lower: the current constraints on the PBH abundance for a monochromatic population (see the main text for a description of the plotted bounds). The colored lines in the upper panel indicate different values of the PBH abundance, while the light blue band in the lower panel shows the number density of observed LRDs associated with an absolute UV magnitude between ${\rm M}_\text{\tiny UV} \approx -17$ and  ${\rm M}_\text{\tiny UV} \approx -22$~\cite{Kocevski:2025tft}.}
    \label{fig:direct}
\end{figure}

\vspace{0.1cm}
\noindent{{\bf{\em (i) Direct PBH formation.}}} 
We examine the possibility that the BHs observed in LRDs could originate from PBHs formed directly with the observed masses in the early Universe, given the existing constraints on the PBH population summarized in Fig.~\ref{fig:direct}. We assume PBH formation proceeds from the collapse of large density perturbations. This typically requires enhanced curvature perturbations $\mathcal{R}$ on small scales~\cite{Blinnikov:2016bxu,Ivanov:1994pa,Ivanov:1997ia,Kawasaki:1997ju}, which can arise in a variety of early Universe scenarios~\cite{Yokoyama:1998pt, Ferrante:2023bgz}, with a primordial power spectrum  $\mathcal{P}_\mathcal{R}$ reaching ${\cal O}(10^{-2})$. Regions exceeding a critical threshold upon Hubble re-entry can collapse to form PBHs across a wide range of masses~\cite{Germani:2018jgr,Musco:2018rwt,Musco:2020jjb}. Within this scenario, the PBH mass is related to the corresponding comoving wavenumber $k$~\cite{Franciolini:2022tfm} (see note in ~\cite{kappa}):
\begin{equation}
M_\text{\tiny PBH} \simeq  
 44 \, M_\odot
 \left(
 \frac{10^{6} \, {\rm cMpc}^{-1}}{k}\right)^{2}\,,
 \label{standardREL}
\end{equation}
showing that directly sourcing PBHs as massive as~$M_\PBH \sim (10^{7} - 10^{8})M_\odot$ would require the gravitational collapse of ${\rm ckpc}$-size perturbations, entering the range of scales constrained by cosmic microwave background (CMB) experiments. In particular, as shown in the upper panel of Fig.~\ref{fig:direct}, the LRD mass range is severely constrained by the FIRAS experiment (brown region), which measures the $\mu$-distortions sourced by the energy injection into the baryon-photon thermal bath from the damping of large amplitude density perturbations~\cite{Chluba:2019nxa}. 
This implies that PBHs cannot have masses larger than $\sim {\cal O}(10^3) M_\odot$ (see Fig.~\ref{fig:direct}), unless the perturbations
were extremely non-Gaussian, an (in principle) viable possibility (see e.g.~\cite{Nakama:2017xvq,Iovino:2024tyg}) that is not easily realized in realistic scenarios~\cite{Unal:2020mts,Byrnes:2021jka,Juan:2022mir,Hooper:2023nnl,Byrnes:2024vjt,Pritchard:2025yda}. Slightly lighter PBHs are also constrained by pulsar timing arrays (PTA) data (magenta region)~\cite{NANOGrav:2023hvm,NANOGrav:2023hfp,NANOGrav:2023gor,EPTA:2023fyk,EPTA:2023sfo,EPTA:2023xxk, Reardon:2023gzh,Zic:2023gta,Reardon:2023zen, Xu:2023wog}, given that sizable perturbations would source a scalar-induced stochastic background~\cite{Iovino:2024tyg, Franciolini:2023pbf,Inomata:2023zup,Cecchini:2025oks}.

The probability $\beta(M_\PBH)$ to form a PBH of mass $M_\PBH$
is extremely sensitive to the statistics of rare, large fluctuations in the density field.
The probability of collapse dictates the fraction of DM contained in PBHs $f_{\textrm{\tiny PBH}}$ as~\cite{Sasaki:2018dmp}
\begin{equation}
\label{beta}
f_{\textrm{\tiny PBH}}(M_\PBH) \simeq
\left(\frac{\beta(M_\PBH)}{6 \cdot 10^{-9}}\right) \left(\frac{M_\PBH}{M_\odot}\right)^{-1/2}\,.
\end{equation}
In the upper panel of Fig.~\ref{fig:direct}, we show the amplitude of ${\cal P}_{\cal R}$ required to reproduce selected values of $f_\text{\tiny PBH}$. We follow state-of-the-art modeling of the collapse probability (see, e.g., Ref.~\cite{Cecchini:2025oks} and the publicly available code~\cite{fastPTAmodule}). We conservatively report the computation adopting threshold statistics~\cite{Musco2025}, but acknowledge that systematic uncertainties remain, as peak-theory computations~\cite{Bardeen:1985tr} indicate that lower amplitudes are needed and, thus, much weaker bounds~\cite{Young:2014ana,DeLuca:2019qsy,Iovino:2024tyg,Cecchini:2025oks}. This, however, does not affect our conclusions. 
For the role of non-Gaussianities in the computation of $f_\PBH$, see~\cite{Yoo:2018kvb,Byrnes:2012yx,Atal:2018neu, DeLuca:2022rfz, Franciolini:2018vbk,Ferrante:2022mui,DeLuca:2019qsy,Young:2019yug,Biagetti:2021eep,Gow:2022jfb}.

Once formed, PBHs are expected to follow a spatial distribution determined by the properties of the underlying density perturbations. For Gaussian fields, PBHs are essentially uncorrelated and their distribution is  described by a rare Poisson process~\cite{Desjacques:2018wuu,Ali-Haimoud:2018dau,Ballesteros:2018swv,MoradinezhadDizgah:2019wjf} (in the presence of primordial non-Gaussianity, PBHs can exhibit spatial correlations~\cite{Atal:2020igj,DeLuca:2021hcf, DeLuca:2022uvz, DeLuca:2022bjs}, which are, however, difficult to achieve in standard scenarios~\cite{Crescimbeni:2025ywm}). The mean comoving number density of a monochromatic PBH population  is
\begin{equation}
n_\PBH \simeq 3 \, \left(\frac{f_\PBH}{10^{-7}} \right)\left(\frac{M_\PBH}{10^3 M_\odot}\right)^{-1} \textrm{cMpc}^{-3}\,.
\end{equation}

 The PBH abundance is constrained by various probes~\cite{Carr:2020gox}, see lower panel of Fig.~\ref{fig:direct}. Between $\sim 10^{-11}\,M_\odot$ to a few tens of $M_\odot$, microlensing surveys (M, blue contour) limit PBHs through the non-detection of lensing events~\cite{Niikura:2017zjd,Niikura:2019kqi,Mroz:2024wag,Mroz:2024wia}, while around stellar masses, gravitational-wave detections constrain PBHs by comparing observed merger rates with PBH merger predictions~\cite{LIGOScientific:2019kan,Kavanagh:2018ggo,Wong:2020yig,Hutsi:2020sol,DeLuca:2021wjr,Franciolini:2022tfm,Andres-Carcasona:2024wqk} (not shown in the plot to avoid cluttering). Above the solar mass scale, CMB~\cite{Serpico:2020ehh,Agius:2024ecw,Manshanden:2018tze,Jensen:2024hpd} (and x-ray backgrounds~\cite{Casanueva-Villarreal:2025kmd, Ziparo:2022fnc}) would be affected by baryonic accretion, depending on the accretion model, e.g. Bondi-Hoyle-Lyttleton (BHL, red contour)~\cite{Serpico:2020ehh} or Park-Ricotti (PR, orange contour)~\cite{Park:2010yh,Park:2011rf,Park:2012cr,Sugimura:2020rdw,Scarcella:2020ssk}. 
For $M_\PBH \gtrsim 10^5\,M_\odot$, constraints arise from the disruption of wide binaries, globular cluster stability, and dwarf galaxy survival (DF, green contour)~\cite{Carr:2018rid,Carr:2020erq}, and large-scale structure formation (LSS, magenta contour)~\cite{Carr:2018rid,Carr:2020erq}. 
Further constraints dominating at masses $M_\text{\tiny PBH} \gtrsim 10^9 M_\odot$ arise from CMB bounds on isocurvature perturbations \cite{Gerlach:2025vco}, not shown for simplicity. Finally, the brown  constraint shows the region excluded by $\mu$-distortions discussed above~\cite{Chluba:2019nxa}.

The combined cosmological constraints shown in Fig.~\ref{fig:direct} demonstrate that the direct formation of PBHs with masses in the range~$ \sim (10^{4} - 10^{11})M_\odot$ is ruled out. Generating PBHs of this size would require a curvature power spectrum amplitude that is incompatible with FIRAS limits on CMB $\mu$-distortions. Consequently, the permitted ${\cal P}_{\cal R}$ yields only a negligibly small abundance of PBHs in this mass window (as illustrated by the colored lines in the upper panel of the figure), excluding this proposed interpretation of LRDs. We stress that masses~$\sim 10^3 M_\odot$ provide the heaviest seeds allowed by cosmology, for which the power spectrum can reach values $\sim 10^{-2}$, compatible with sizable PBH production. As shown in the lower panel, PBHs with these masses can comprise at most $f_\PBH \sim (10^{-4} - 10^{-7})$ of the DM, and may only act as seeds to grow into heavier BHs through hierarchical mergers and baryonic accretion, as we discuss below. 
We stress that the bounds in Fig.~\ref{fig:direct} are redshift-dependent~\cite{Carr:2020xqk} and apply to the initial PBH masses (note that substantial PBH accretion shifts observational bounds to larger masses~\cite{DeLuca:2020fpg}). This implies that PBHs with initial masses  $\sim 10^3\,M_\odot$ can, in principle, give rise to a sizable population of LRDs through cosmological growth (following the lines of constant number density, blue arrows), without violating the $\mu$-distortion limits, which are relevant only for larger initial masses.

\vspace{0.1cm}
\noindent{{\bf{\em (ii) Assembly via hierarchical mergers.}}} 
As lighter PBHs remain viable, we now turn to studying the growth of these seeds through hierarchical mergers. 
These represent a growth mechanism in which PBHs form binaries, merge, and then continue to coalesce with other PBHs in subsequent generations, thus building up mass~\cite{Yang:2019cbr, Antonini:2015zsa, Antonini:2018auk}. This growth can proceed through two main channels: a democratic one, in which PBHs of comparable masses merge pairwise at each step, and an oligarchic one, in which the most massive remnants produced in each generation repeatedly merge with lighter seeds~\cite{Kovetz:2018vly}.
The true scenario is therefore expected to lie between these two limiting cases.
A crucial condition for this to work is that the mergers occur within environments dense enough to retain the remnants of each generation. Since these remnants receive recoil kicks~\cite{Varma:2020nbm}, 
only sufficiently deep potential wells can keep them bound, allowing them to participate in the hierarchical growth. 

To provide a concrete example, we investigate the possibility of reproducing the properties of the event QSO1. We stress that our methodologies are general and can be applied to any LRDs. Concerning QSO1, given the low stellar content observed, we expect that the only relevant environment where a hierarchical assembly can be obtained is within DM halos. In these systems, $N_\PBH$ PBHs are expected to undergo mass segregation and become concentrated within the radius~\cite{2013MNRAS.432.2779B}
\begin{equation}
r_\text{\tiny seg} \simeq \frac{10}{N_\PBH^{3/2} \log^{1/3}(0.02 N_\PBH)}  R_\text{\tiny 200}\,,
\end{equation}
in terms of the radius $R_\text{\tiny 200}$ at which the halo mean density is 200 times the critical density of the Universe.
For such environments, retaining merger remnants requires kick velocities smaller than the escape velocity, $v_\text{\tiny esc} (r_\text{\tiny seg}) = \sqrt{2 |\Phi(r_\text{\tiny seg})- \Phi(2 R_\text{\tiny 200})|}$~\cite{Roche:2024gcl}.
Here, $\Phi$ describes the DM potential, assumed to follow a Navarro-Frenk-White (NFW) profile~\cite{Navarro:1995iw}, with concentration $c_\text{\tiny DM}$.
To reach $ \sim (10^{7} -10^{8})M_\odot$, we would need at least $N_\PBH \sim 10^4$ PBHs in the DM halo. As shown below, considering a more abundant number of PBHs would require more massive DM halos, which are rare at high redshifts. 

We estimate the probability of forming a LRD-like system with mass $10^7 M_\odot$ through the runaway mergers of $N_\PBH \simeq 10^4$ PBHs with mass $M_\PBH \simeq 10^3 M_\odot$, by performing simulations as described in Supplemental Material.
These simulations determine the probability $\epsilon_\text{\tiny HM} (v_\text{\tiny esc})$ of getting a final BH with mass compatible with $10^7 M_\odot$ by varying the escape velocity of the DM environment.
The efficiency can be fitted as
\begin{equation}
\label{fitepsilon}
\epsilon_\text{\tiny HM}(v_\text{\tiny esc}) = \frac{1}{1 + \exp\left[-a\left(\log_{10} v_\text{\tiny esc} - b\right)\right]}\,,
\end{equation}
 with parameters $
a=9.36 \, (17.00)$ and $b=2.99 \, (3.29)$, for the oligarchic (democratic) scenario.

\begin{figure}
    \centering
    \includegraphics[width=\linewidth]{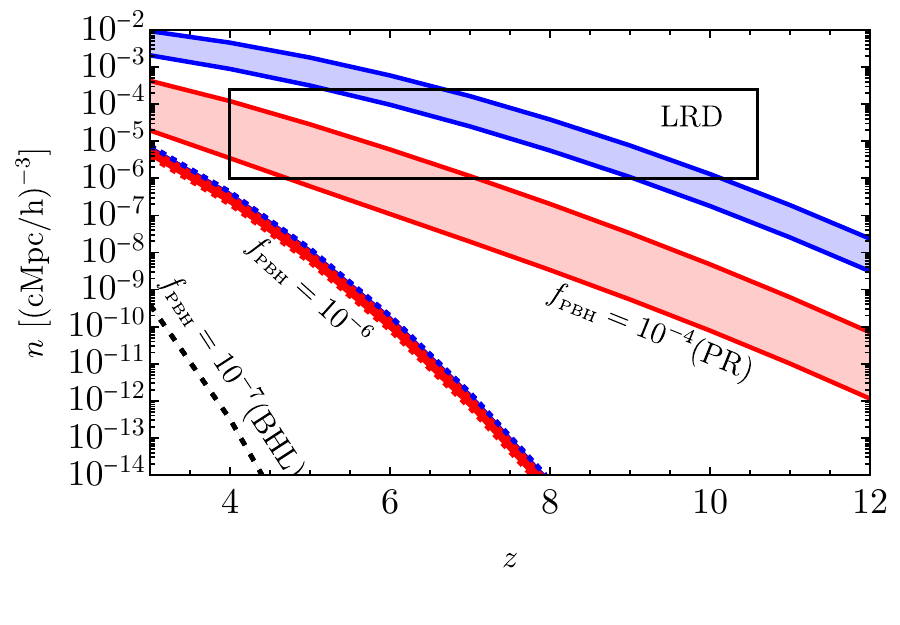}
    \caption{Number density of environments that are able to form a $10^7 M_\odot$ PBH through hierarchical mergers of lighter $10^3 M_\odot$ seeds, as a function of redshifts $z$. Red and blue colors correspond to mergers proceeding through the democratic and oligarchic scenarios, respectively. Solid, dashed and dotted (shown in black as both cases overlap) lines indicate different values of the PBH abundance $f_\PBH$, while the bands consider values of the NFW concentration parameter within the range $c_\text{\tiny DM} = (0.1 - 1)$. The black rectangle indicates the parameter space of the LRDs observed so far~\cite{Dayal:2024zwq}. BHL and PR correspond to assuming the maximum abundance allowed by those constraints at the assumed initial mass $10^3 M_\odot$, see Fig.~\ref{fig:direct}.}
    \label{fig:HM}
\end{figure}

By determining $v_\text{\tiny esc}$ in terms of the DM halo mass $M_h$ (obtained by integrating the NFW profile), we provide the probability $\epsilon_\text{\tiny HM}(M_h)$ that a given DM halo retains the merger remnants and produces a heavier PBH  compatible with QSO1.
This has to be combined with the abundance of DM halos at a given redshift, captured in the halo mass function 
${{\rm d}n(z)}/{{\rm d} M_h}$~\cite{Dodelson:2003ft}. 
The number density of environments able to source QSO1-like events reads
\begin{equation}
n (z)  = \int{\rm d} M_h \frac{{\rm d}n}{{\rm d} M_h} \epsilon_\text{\tiny HM}(M_h)\,.
\end{equation}
Its time evolution is shown in Fig.~\ref{fig:HM}, which summarizes the main predictions for PBH hierarchical mergers. First, one can observe that smaller abundances strongly suppress the probability of sourcing these events. This happens because smaller values of $f_\PBH$ require heavier DM halos to have at least $N_\PBH = 10^4$ objects to undergo the runaway series. Despite the higher escape velocity of these systems (thus retaining each merger remnants), the density of very massive halos decreases at high redshifts, thereby explaining the trends of the plot.
Furthermore, by comparing the oligarchic (blue curves) with the democratic (red curves) scenarios, one appreciates that the former gives rise to a larger probability. This is because remnant kicks are suppressed for larger mass ratios, thus reducing the probability of being ejected. 

The prediction of our model is comparable with the LRDs observed number density, shown in the black rectangle, only in the oligarchic scenario and for values of the PBH abundance as large as $f_\PBH \sim 10^{-4}$, which corresponds to the more optimistic version of the CMB bound (see orange and red contours of Fig.~\ref{fig:direct}), while strongly suppressed otherwise. This implies that explaining the abundance of LRDs and events such as QSO1, using hierarchical mergers alone, is challenging.

\vspace{0.1cm}
\noindent{{\bf{\em (iii) Cosmological growth through accretion.}}} 
The final scenario we consider is one in which the initial PBH seeds with masses $\sim 10^3 M_\odot$ grow through baryonic accretion (see also Ref.~\cite{Volonteri:2025iit} for a recent review). During their cosmological history, PBHs may efficiently accrete baryons from the surrounding medium, potentially increasing their masses, and evolving their spins due to disk formation~\cite{Ricotti:2007au,Ali-Haimoud:2016mbv, DeLuca:2020qqa, DeLuca:2023bcr}, despite several uncertainties related to mechanical and radiative feedback effects~\cite{Ricotti:2007au,Ali-Haimoud:2016mbv, Bosch-Ramon:2020pcz, Facchinetti:2022kbg}.
Since baryonic accretion is inherently complex and achieving a complete understanding generally requires sophisticated simulations, in this Letter we take a deliberately simpler route and adopt an accretion model inspired by Refs.~\cite{DeLuca:2023bcr,Serpico:2024cdz,Jangra:2024sif} (see especially discussion around Fig.~4 of Ref.~\cite{Inayoshi:2019fun}). As in the previous section, we aim to reproduce the physical characteristics of QSO1. Nonetheless, the model we consider is general and can be applied to investigate the accretion evolution of PBHs and their ability to reproduce LRDs. 

Our starting point is a PBH seed of mass $\sim 10^3 M_\odot$, which begins its accretion evolution at high redshifts ($z\sim 1000$ for definiteness). At higher redshifts, the accretion timescale is larger then the age of the Universe, rendering accretion essentially ineffective~\cite{Jangra:2024sif}. At smaller redshifts, PBHs can accrete baryons from the surrounding medium, potentially changing the metallicity content of its neighborhood due to disk fragmentation and star formation. Considering these phenomena, the PBH accretion rate $\dot{M}_\text{\tiny PBH}$ is determined  through the balance law
\begin{align}
\label{balanceM}
 \dot{M}_\text{\tiny PBH}  = \dot{M}_\text{\tiny inf} -  \dot{M}_* - \dot{M}_\text{\tiny out} - \dot{M}_\text{\tiny g}\,.
\end{align}
The contribution $\dot{M}_\text{\tiny inf}$ accounts for the infall of matter through the PBH Bondi radius, defined as $r_\text{\tiny B} = {G M_h}/{v_\text{\tiny eff}^2}$~\cite{Ricotti:2007jk}. This radius depends on the effective PBH velocity $v_\text{\tiny eff} = \sqrt{c_s^2 + v_\text{\tiny rel}^2}$,
expressed in terms of the baryonic sound speed $c_s$ and the PBH-baryon relative velocity $v_\text{\tiny rel}$~\cite{Ricotti:2007au,Serpico:2020ehh,Ali-Haimoud:2016mbv,Poulin:2017bwe,Hasinger:2020ptw,Hutsi:2019hlw,Ali-Haimoud:2017rtz},
as well as on the mass enclosed within the surrounding DM minihalo, $M_h = 3 M_\PBH \left( \frac{1000}{1+z} \right)$, which embeds heavy PBHs when they constitute only a subdominant DM component~\cite{Ricotti:2007jk}, as in our case. The infalling rate takes the form
\begin{align}
\dot{M}_\text{\tiny inf} & = 
\begin{cases}
\dot{M}_\text{\tiny B} \quad z \gtrsim z_\text{\tiny B-inst}\,, \\
\frac{c_s^3}{G} \quad z \lesssim z_\text{\tiny B-inst}\,. \\
\end{cases}
\end{align}
The first term describes the spherically symmetric Bondi accretion rate, $\dot{M}_\text{\tiny B} = 4 \pi \lambda \rho \,r_\text{\tiny B}^2 v_\text{\tiny eff}$~\cite{1939PCPS...35..405H, Bondi:1944rnk}, 
 onto a mass embedded in a uniform medium with density $\rho$ and effective accretion parameter $\lambda$~\cite{Ricotti:2007au, Ricotti:2007jk}, which phenomenologically captures the influence of non-gravitational effects that reduce the accretion rate, including the Hubble expansion, the coupling of the CMB radiation to the gas through Compton scattering, and the gas viscosity. In what follows, we treat $\lambda$ as a free parameter and constrain its allowed range by requiring consistency with the observed LRD properties.
 This contribution provides the flow of matter onto the PBH up to redshift $z_\text{\tiny B-inst}$, which is set by the Bondi instability time scale~\cite{Alexander_2014}, at which the gas surrounding the PBHs becomes ionized along the polar axis because of radiation feedback~\cite{Ricotti:2007au,Ali-Haimoud:2016mbv, Facchinetti:2022kbg} (at variance with the Park-Ricotti model, where the gas within the Bondi radius is completely ionized~\cite{Park:2010yh,Park:2011rf,Park:2012cr,Sugimura:2020rdw,Scarcella:2020ssk}). At this point, the sound speed within the Bondi radius increases due to ionization, reaching values $c_s \approx 10 \ {\rm km/s}$,  thus reducing the accretion rate to $c_s^3/G$~\cite{Inayoshi:2019fun}.

The second contribution in Eq.~\eqref{balanceM}, $\dot{M}_*$, comes from the possibility that, as efficiently accreting PBHs are expected to develop an accretion disk surrounding them~\cite{DeLuca:2020bjf,DeLuca:2020qqa,DeLuca:2023bcr}, such disks can effectively fragment, 
inducing star formation~\cite{Toomre:1964zx}. This process is usually described in terms of the  Toomre parameter, $Q  \sim \mathcal{O}(1) c_s^3/G \dot{M}_\text{\tiny inf}$, so that regions with $Q \lesssim 1$ are susceptible to gravitational collapse~\cite{Goodman:2002gv}. This implies that, without rapid inward migration of fragments~\cite{Inayoshi:2014ija}, star formation may already occur at the onset of the Bondi instability, when the density within the Bondi sphere rises significantly. We introduce the effective parameter $f_\text{\tiny frag} \equiv \dot{M}_*/\dot{M}_\text{\tiny inf}$ to track the fraction of incoming flow which fragments. 

Third, in Eq.~\eqref{balanceM}, we account for the ejection of a fraction of the incoming matter by adding $\dot{M}_\text{\tiny out} \equiv f_\text{\tiny out}\dot{M}_\text{\tiny inf}$. 

Finally, we introduce a contribution, $\dot{M}_\text{\tiny g}$, accounting for
gas build-up within the Bondi sphere when the system is not in a steady state.
This component is expected to be small before the instability~\cite{Dayal:2025aiv}, when most of the inflowing matter is efficiently accreted by the PBH. However, at  $z_\text{\tiny B-inst}$, we expect a significant gas mass within the region, $M_\text{\tiny g} \sim M_\PBH (z_\text{\tiny B-inst})$. After the instability time, we assume a steady-state configuration with constant gas mass. While it is theoretically possible to consider a growth in gas mass proportional to the infall rate, $\dot{M}_\text{\tiny g} \propto \dot{M}_\text{\tiny inf}$, this effect is degenerate with changes in the ejection fraction, $f_\text{\tiny out}$, and thus is omitted in our Letter.

 \begin{figure}
    \centering
    \includegraphics[width=1
    \linewidth]{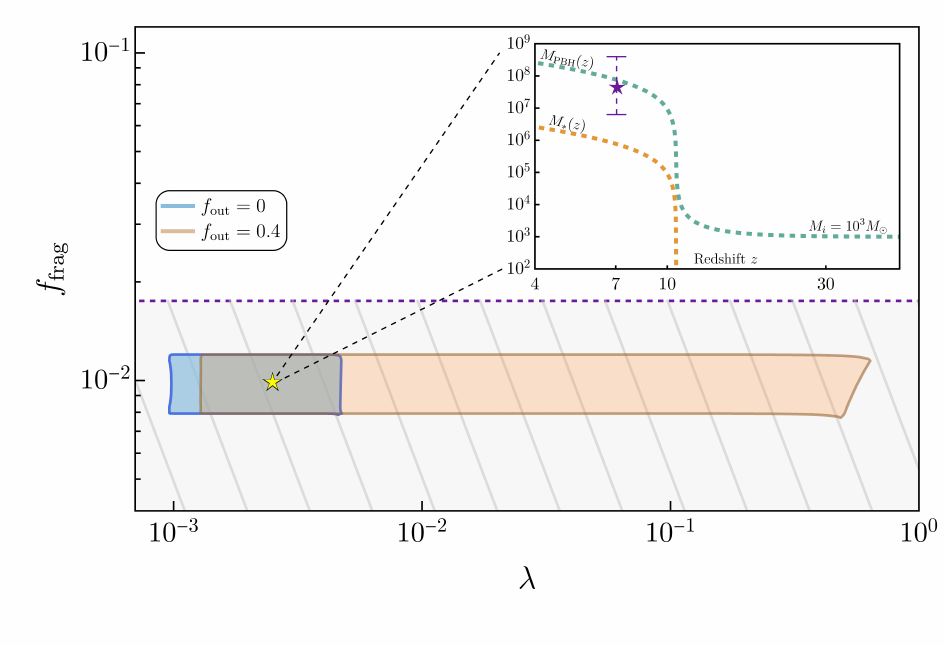}
    \caption{Parameter space of the accretion-metallicity model, defined by the fragmentation and accretion coefficients,  $f_\text{\tiny frag}$ and $\lambda$, required to reproduce QSO1 within its $1\sigma$ confidence intervals, $M_\text{\tiny QSO1} = 10^{7.7} \, M_\odot$ and $Z_\text{\tiny QSO1} = 10^{-2.08}Z_\odot$. The orange and blue contours  correspond to different choices of the ejection parameter $f_\text{\tiny out}$. The inset shows the redshift evolution of the PBH and stellar masses corresponding to the yellow star, which marks the central values of the QSO1 parameters. The dashed purple line indicates the maximum $f_{\text{\tiny frag}}$ allowed within $3\sigma$ of the observed metallicity. The gray region shows the parameter space where the metal enrichment from structure formation  could contribute to the metallicity of QSO1.
    }
    \label{fig:accretion}
\end{figure}

Together with mass evolution, the presence of star formation induces a change in the metallicity $Z$ of the system. Similar to Eq.~\eqref{balanceM}, one can write an effective balance law for this quantity, which leads to~\cite{Dayal:2025aiv}
\begin{equation}
\label{balanceZ}
\dot{Z}  = \frac{1}{M_\text{\tiny g}}(y f_\text{\tiny frag} - Z ) \dot{M}_\text{\tiny inf}\,,
\end{equation}
in terms of the net metal yield $y \sim \mathcal{O}(10^{-2})$~\cite{Madau:2014bja}. As expected, a larger fragmentation parameter would result in a larger metal enrichment of the environment. Notice that our modeling neglects the contribution of heavy elements, associated with the usual structure formation history at later redshifts~\cite{Madau:2014bja}. In this sense, our results have to be intended as an upper bound on the fragmentation fraction $f_\text{\tiny frag}$ of QSO1-like LRDs (see gray region in Fig.~\ref{fig:accretion}). 

In Fig.~\ref{fig:accretion} we show the parameter space of our model aimed at reproducing the features of QSO1~\cite{Maiolino:2025tih, 2025arXiv250821748J}. The contours indicate that the accretion parameter can span the range $\lambda \in (10^{-3} - 1)$ and effectively account for the mass growth of PBHs from $10^3\,M_\odot$ to $10^7\,M_\odot$. Notably, increasing the fraction $f_\text{\tiny out}$  of ejected matter (orange contour) effectively corresponds to considering a larger accretion parameter. For the relevant range of $\lambda$, we numerically verified that the dimensionless accretion rate $\dot{m} \equiv  \dot{M}_{\text{\tiny PBH}} / \dot{M}_{\text{\tiny Edd}} \lesssim 1$ for $ z\gtrsim z_\text{\tiny B-inst}$, where $\dot{M}_{\text{\tiny Edd}} = 10 \,L_{\text{\tiny Edd}}$~\cite{Xie_2012}. For $z\approx z_\text{\tiny B-inst}$, we enter a super-Eddington phase where $\dot{m}$ reaches values as high as $\sim 10^2$. Depending on the value of $\lambda$, the BH may still be in a super-Eddington accretion phase at redshifts $z \approx 7$. Within such a phase, the radiation efficiency is expected to decrease significantly due to the expansion of the trapping radius~\cite{Zhang:2025uug}, resulting in the absence of x-ray and variability signatures~\cite{Lei:2008ui,Madau:2025kpn}, as observed for LRDs~\cite{Pacucci:2024tws, Lambrides:2024ugh, Madau:2024fdv,Inayoshi:2025lsf}. 
Furthermore, the corresponding luminosity is consistent with the observed value, $L_{\text{\tiny QSO1}} \simeq 10^{-2} L_{\text{\tiny Edd}}$, assuming an accretion efficiency of order~$\mathcal{O}(10^{-3} - 10^{-4})$. This value aligns closely with predictions from the slim disk model~\cite{Abramowicz:1988sp,Zhang:2025uug}.
Conversely, the fragmentation coefficient $f_\text{\tiny frag}$ primarily controls the metal content around the PBH. Since QSO1 exhibits exceptionally low metallicity, this sets an upper limit $f_\text{\tiny frag} \lesssim 10^{-2}$ to avoid overproducing heavy elements. Overall, our model demonstrates that, despite uncertainties in the accretion physics, PBHs could plausibly reproduce the observed properties of QSO1. Our upper limit on $f_\text{\tiny frag}$ implies that the produced stellar mass remains much smaller than the PBH mass, $M_\star \lesssim 10^{-2} M_\PBH$ (see inset in Fig.~\ref{fig:accretion}), consistent with the inferred ratio $M_\text{\tiny BH}/M_\star \gtrsim 2$. Such a large PBH-to-star mass ratio may be a smoking gun of the primordial interpretation for QSO1-like events.

\vspace{0.1cm}
\noindent{{\bf{\em Conclusions.}}}
The discovery of LRDs poses a challenge to standard models of early galaxy formation, given their combination of compact sizes, low stellar masses, and potentially massive BHs (we emphasize that measurements of the BH masses based on virial relations may suffer large uncertainties if the observed broad-line widths are dominated by electron scatterings rather than gravitational motions~\cite{Naidu:2025rpo, 2026Natur.649..574R, 2026MNRAS.546ag086J}). In this work, we explored whether LRDs could have a primordial origin, focusing on the properties of QSO1 to streamline our analysis. Remarkably, QSO1 is the first LRD with a dynamically measured BH mass \cite{2025arXiv250821748J} and it exhibits exceptionally low metallicity \cite{Maiolino:2025tih}.

\FloatBarrier
\begin{table}[t!]
\centering
\renewcommand{\arraystretch}{1.2}
\begin{tabular}{|l|c|l|}
\hline
Scenario &  & Constraint \\
\hline
{\it (i)} Primordial direct collapse & \xmark & $\mu$-distortions\\
\hline
{\it (ii)} Hierarchical assembly & \xmark & Low number density \\
\hline
{\it (iii)} Cosmological accretion & \cmark & Low fragmentation needed\\
\hline
\end{tabular}
\caption{Summary of the primordial scenarios considered in this Letter.}
\label{tab:flowchart}
\end{table}

We have examined three PBH-based pathways for the origin of the BHs powering LRDs, briefly summarized in Table~\ref{tab:flowchart}.  
We have shown that the direct formation of such massive objects as PBHs is ruled out by CMB $\mu$-distortion constraints, which severely limit the abundance of very massive PBHs at early times. This implies that scenarios based on PBH seeds as heavy as $10^5\,M_\odot$, as in Refs.~\cite{Zhang:2025asq,Maiolino:2025tih}, are excluded by cosmological bounds. Consequently, viable PBH seeds are restricted to masses $\lesssim 10^3\,M_\odot$, constituting only a subdominant component of the total DM. Hierarchical mergers of lighter, observationally allowed PBHs provide a possible growth channel, but the efficiency of this channel is highly sensitive to the assembly history and constrained by the typical properties of high-redshift DM halos, making it challenging to reproduce the observed masses. By contrast, gas accretion onto intermediate-mass PBHs, when modeled self-consistently with metallicity evolution, emerges as a more robust pathway, identifying regions of parameter space that can simultaneously reproduce the observed BH masses, stellar-to-BH ratios, and metallicities of LRDs,  including the properties of the event QSO1. These conclusions are conservative because PBH growth can happen through a combination of accretion and mergers.

Our work can be extended in several directions. One possibility concerns applying this analysis to the full LRD population, assessing whether PBH-based channels can consistently reproduce their observed properties. 
This could include, among others, the highest-redshift source reported to date~\cite{2025ApJ...995...21T} or candidate dual LRDs~\cite{2024arXiv241214246T}.
Moreover, refining the accretion model will be essential to better constrain the parameter space of viable PBH scenarios. 
Additionally, it would be interesting to consider a mixture scenario in which both accretion and merger contribute to the population evolution. Finally, one may also consider initially clustered PBH formation (e.g.~\cite{DeLuca:2022bjs,Zhang:2025tgm}), a scenario which also requires a careful assessment of the CMB bounds as discussed in \cite{DeLuca:2021hcf}. We leave these considerations to future work. 

\noindent
{\bf Note added:}
Recently, Ref.~\cite{Zhang:2025oyl} appeared, showing that the accretion and metallicity evolution of a heavy PBH seed can result in a QSO1-like LRD when considering both BH and stellar feedback. Unlike in our Letter, the authors assume the PBH seed to be formed directly with a large observed mass.

\vspace{0.1cm}
\noindent{{\bf{\em Acknowledgments.}}}
We thank P.~Dayal and R.~Maiolino for interesting discussions.
E.B., V.D.L., L.D.G and K.K. are supported by NSF Grants No.~AST-2307146, No.~PHY-2513337, No.~PHY-090003, and No.~PHY-20043, by NASA Grant No.~21-ATP21-0010, by John Templeton Foundation Grant No.~62840, by the Simons Foundation [MPS-SIP-00001698, E.B.], by the Simons Foundation International [SFI-MPS-BH-00012593-02], and by Italian Ministry of Foreign Affairs and International Cooperation Grant No.~PGR01167.
K.K. is supported by the Onassis Foundation  - Scholarship ID: F ZT 041-
1/2023-2024. This work was carried out at the Advanced Research Computing at Hopkins (ARCH) core facility (\url{https://www.arch.jhu.edu/}), which is supported by the NSF Grant No.~OAC-1920103.
G.F.~acknowledges support by the
Italian MUR Departments of Excellence grant 2023–2027
“Quantum Frontier” and from Istituto Nazionale di Fisica Nucleare (INFN) through the Theoretical Astroparticle Physics (TAsP) project.

\bibliography{draft}

 \clearpage

\appendix
 \onecolumngrid
\begin{center}
    {\bf Supplemental Material}
\end{center}
\section{More details on the hierarchical formation scenario} \label{app:hierarchical}
\noindent
In this appendix we provide additional details on the hierarchical merger scenario discussed in the main text, and quantify the probability that a purely hierarchical growth process can assemble the $10^7\,M_\odot$ BH associated with QSO1. 

We estimate the probability of forming such a BH through a runaway process starting from PBHs of mass $M_\PBH=10^3\,M_\odot$. As discussed in the main text, we consider either a democratic growth scenario, in which the BH mass doubles at each merger, or an oligarchic case, in which a single runaway proceeds by repeatedly merging with single $10^3\,M_\odot$ PBHs. These models bracket the lower and upper bounds of the hierarchical merger efficiency $\epsilon_\text{\tiny HM}(v_\text{\tiny esc})$, defined as the probability that the merger chain reaches $10^7\,M_\odot$, without ejecting the runaway remnant, for a given escape velocity $v_\text{\tiny esc}$ of the DM environment within which these objects reside. Defining $N_\text{\tiny m}$ as the number of mergers in each scenario, a successful sequence requires either $10^3\,M_\odot\cdot 2^{N_\text{\tiny m}}=10^7\,M_\odot$ (democratic) or $10^3\,M_\odot \cdot N_\text{\tiny m}=10^7\,M_\odot$ (oligarchic), implying in both cases a minimum cluster size of about $N_\text{\tiny PBH}\simeq10^4$ PBHs. While increasing the number of objects would facilitate mergers in dense environments, such configurations are statistically disfavored by current observational bounds on the PBH abundance, as we discussed in the main text. 

Within these simulations, we assume an exponential runaway time growth and a constant escape velocity. PBHs are assumed to be nonspinning at formation~\cite{DeLuca:2019buf, Mirbabayi:2019uph}, but we consistently track their spin growth during the hierarchical evolution (we assume all BHs have the same spin magnitudes after each merger and randomize over orientations in the democratic scenario). 
Furthermore, we compute the merger remnant properties using numerical relativity fitting formulae~\cite{Varma:2020nbm,Varma:2019csw, Islam:2025drw}, and terminate the hierarchical merger process whenever the gravitational wave recoil exceeds the escape velocity. This model is optimistic, as it neglects the dynamical friction return timescale of slightly kicked runaway objects not completely ejected from the halo. We also neglect three-body ejection mechanisms, which could enhance the removal rate of BHs from dense environments, thereby suppressing hierarchical growth~\cite{DallAmico:2023neb}. This effect would make our argument even stronger. Our results are averaged over 100 realizations for 50 selected values of $v_\text{\tiny esc}$.

\begin{figure}[h] \centering \includegraphics[width=0.6\linewidth]{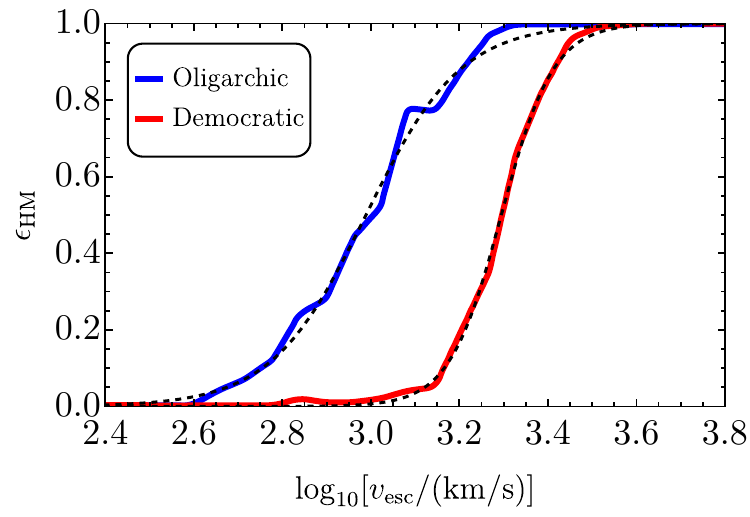} \caption{Behavior of the hierarchical merger probability as a function of the environment escape velocity. Solid lines show the numerical results, while dashed lines indicate the analytical fit using the sigmoid function presented in Eq.~\eqref{fitepsilon} of the main text.} \label{fig:HM-efficiency}\end{figure}

The resulting behavior of the hierarchical merger probability is shown in Fig.~\ref{fig:HM-efficiency} for the oligarchic (blue) and democratic (red) scenarios, as a function of the halo escape velocity. The solid lines indicate the results of the numerical simulations, while the dashed lines indicate the analytical fit provided in Eq.~\eqref{fitepsilon} of the main text. We determine the slope parameters $(a,b)$ by fitting the exponential analytical form to the numerically computed merger efficiency using a nonlinear least-squares minimization based on the Levenberg--Marquardt algorithm. We obtain
$a = 9.36\,(17.00)$ and $b = 2.99\,(3.29)$
for the oligarchic (democratic) scenario. As expected, larger escape velocities lead to a higher probability, since gravitational kicks are less likely to eject merger remnants. Similarly, the oligarchic scenario yields a higher probability because, for a fixed escape velocity, the more unequal mass ratios result in smaller recoil kicks for the merger remnants.
Our results remain essentially unchanged when we consider a slightly heavier final BH mass of $10^8\,M_\odot$.

\begin{figure}[h] \centering \includegraphics[width=0.6\linewidth]{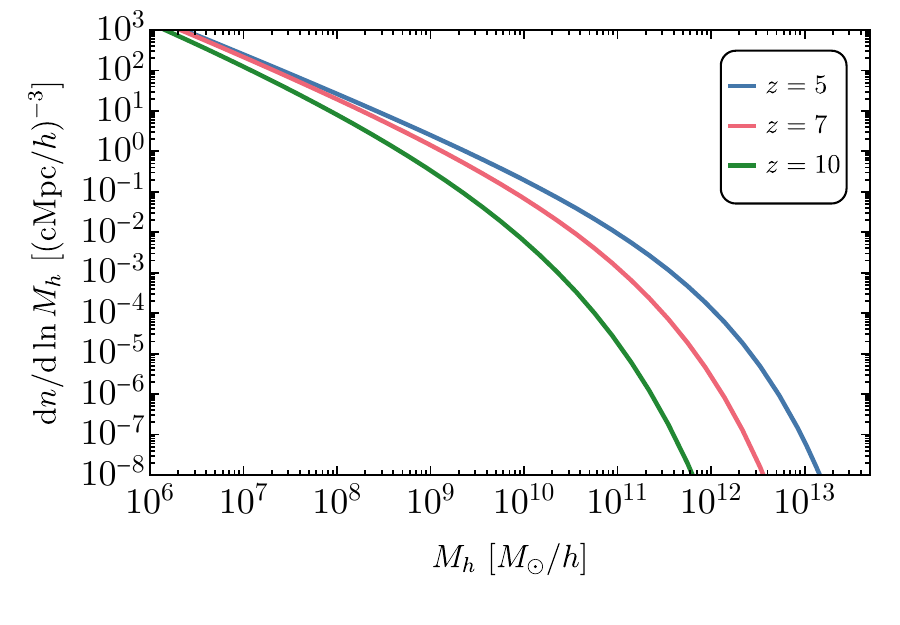} \caption{Halo mass function at various redshifts of interest for this work. } \label{fig:HM-PS}\end{figure}

As discussed in the main text, the probability of efficiently assembling LRD-like objects also depends on the statistical properties of environments that can effectively retain merger remnants. For DM halos, this statistical description is typically formulated using the Press-Schechter formalism, which yields the following expression for the halo mass function~\cite{Dodelson:2003ft}:
\begin{equation}
    \frac{{\rm d} n}{{\rm d}M_h} = F(\nu) \frac{\rho_\text{\tiny m}}{M_h^2} \frac{{\rm d} \ln \sigma^{-1}}{{\rm d} \ln M_h}\,. 
\end{equation}
The mass function depends on the background average matter density $\rho_\text{\tiny m}$, the function $F (\nu)$~\cite{2010ApJ...724..878T} (tested with numerical simulations in~\cite{Biagetti:2022ode} for the masses and redshifts considered here)  with $\nu = \delta_c/\sigma$, expressed in terms of the critical linear overdensity for collapse, $\delta_c = 1.686$, and the variance of the smoothed density field 
\begin{equation}
    \sigma^2 (M_h,z) = \int \frac{{\rm d}^3 k}{(2\pi)^3} W^2(kR) \mathcal{M}^2(k,z) \mathcal{P}_\mathcal{R}(k)\,.
\end{equation}
Such variance is built out of linear perturbations, properly smoothed on the  scale $R = \left( \frac{3M_h}{4\pi \rho_\text{\tiny m}} \right)^{1/3}$ through a top-hat spherical window function in Fourier space
\begin{equation}
    W(kR) = \frac{3 \sin(kR)}{(kR)^3} - \frac{\cos(kR)}{(kR)^2}\,, 
\end{equation}
and assumes the time evolution function
\begin{equation}
    \mathcal{M}(k,z) = \frac{2}{5} k^2 T(k) D(z)\,, 
\end{equation}
in terms of the linear growth factor $D(z)$ and  the linear transfer function $T(k)$, following standard conventions in the literature~\cite{Dodelson:2003ft}. We neglect the additional power introduced by PBH shot noise in the matter distribution, which would only be relevant at smaller scales, and anyway subdominant due to the small abundance assumed here.
The halo mass function is shown in Fig.~\ref{fig:HM-PS}, illustrating the exponential suppression of the tail of higher-mass DM halos depending on redshift.

\end{document}